\documentclass{iau} 
\usepackage{graphicx}

\title[BINOCS: Binary Populations in Clusters] %% give here short title %%
{Binary Information from Open Clusters Using SEDS (BINOCS) Project:
  The Dynamical Evolution of the Binary Populations in Cluster Environments}

\author[Peter Frinchaboy \& Benjamin Thompson]   %% give here short author list %%
{Peter Frinchaboy$^1$
 \and Benjamin Thompson$^1$}

\affiliation{$^1$Department of Physics \& Astronomy, Texas Christian
  University, \\ 2800 S. University Ave., Fort Worth, TX, USA 76129\\ email: {\tt
  p.frinchaboy@tcu.edu, b.a.thompson1@tcu.edu}}

\pubyear{2015}
\volume{316}  %% insert here IAU Symposium No.
\jname{Formation, evolution, and survival of massive star clusters}
\editors{C. Charbonnel \& A. Nota, eds.}
\begin{document}

\maketitle

\begin{abstract}
Studying the internal dynamics of stellar clusters is conducted
primarily through N-Body simulations. One of the major inputs into
N-Body simulations is the binary star frequency and mass distribution,
which is currently constrained by relations derived from field binary
stars. However to truly understand how clustered environments evolve,
binary data from within star clusters is needed including
masses. Detailed information on binaries masses, primary and
secondary, in star clusters has been limited to date. The primary
technique currently available has been radial velocity surveys that
are limited in depth. Using previous two-band photometry-based studies
that may cover different mass ranges produce potentially discrepant
interpretations of the observed binary population. We introduce a new
binary detection method, Binary INformation from Open Clusters Using
SEDs (BINOCS) that covers the wide mass range needed to improve
cluster N-body simulation inputs and comparisons. Using newly-observed
multi-wavelength photometric catalogs (0.3 - 8 microns) of the key
open clusters with a range of ages, we can show that the BINOCS method
determines accurate binary component masses for unresolved cluster
binaries through comparison to available RV-based studies. Using this
method, we present results on the dynamical evolution of binaries from
0.4 - 2.5 solar masses within five prototypical clusters, spaning 30
Myr to 3.5 Gyr, and how the binary populations evolve as a function of
mass.
\end{abstract}

\firstsection % if your document starts with a section,
\section{Introduction}

Current cluster binary studies are carried out using one of two
methods, two-band photometry, and time-baseline radial velocity studies, each of which experience issues which limit their effectiveness in answering the above science questions.
However, to deeply understand the binary populations of open clusters,
we have created a new method which can determine accurate masses for all members of a cluster within a reasonable amount of telescope time. 
This new binary detection method is nicknamed
\textsc{binocs}: \textsc{Binary INformation from Open Clusters using
  SEDs}. By imaging a star using multiple filters across the spectrum
(e.g., $UBVRIJHK_S[3.6][4.5][5.8][8.0]$), one should be able
to ``re-build'' \emph{spectral energy distribution} (SED) of a star
given its parameters: age, metallicity, mass. Similarly, a binary
system could not be accurately modelled by a single SED curve, but
instead by two SEDs added together. By matching stars to these models,
mass can be determined, similar to how temperature could be determined
in the idealized blackbody case. Since the star is a member of a cluster with known
parameters, so age and metallicity are given. By matching stars to
models of a library of synthetic single stars SEDs, mass can be
determined. Isochrones often come in coarse mass grids. To overcome this,
stellar parameters and magnitude are cubically interpolated with
respect to mass onto a new mass grid, with a 
spacing of 0.01 M$_{\odot}$ \cite[(Thompson \& Frinchaboy, 2016)]{thompson16}\footnote{Software used in this study can be obtained from: https://github.com/bathompso/BINOCS}.

In this work, we present analysis from nine open clusters (M35, M36,
M37, M67, NGC 188, NGC 2158, NGC 2420, NGC 6791. NGC 6819), a sample designed to cover a wide range of ages and metallicites within the open cluster population, as well as provide good ties to other open cluster binaries studies for verification. The method depends on having high-quality photometry covering a wide range of wavelengths. 

\vspace*{-0.5 cm}
\section{Results: Full Sample BINOCS }
\begin{figure}[h]
 \vspace*{-0.3 cm}
\begin{center}
 \includegraphics[width=4.0in]{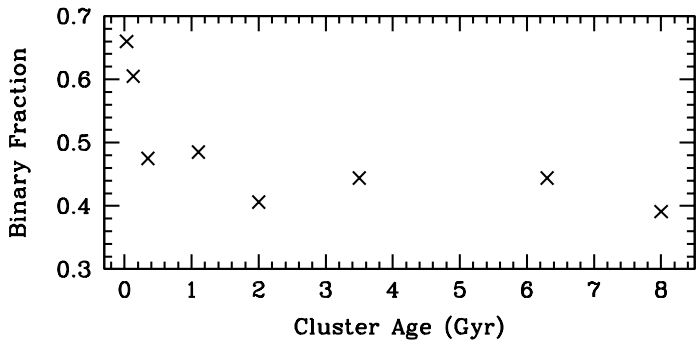} 
 \vspace*{-0.2 cm}
 \caption{Total binary fraction for the BINCOS clusters sample, which
   shows a clear rapid decline in binary fraction within the first 200
   Myr, and then a slow decline thereafter.}
   \label{fig1}
\end{center}
\end{figure}

Using the sample of clusters from the Table below, we find that there
is a significant binary population in young open clusters, with 60-70\% of stars in binary systems.  This significant binary population is quickly disrupted within the first 200 Myr, likely the destruction of wide binaries in the cluster, with a slower rate of disruption thereafter (Fig.\,\ref{fig1}).

\vspace*{-0.7 cm}
\section{Results: Intra-Cluster BINOCS} Binary systems are more
massive, on average, than a single star, and should therefore
experience mass segregation. This has been observationally confirmed
for several globular and open clusters (e.g., \cite[Geller \& Mathieu 2012]{geller12},
\cite[Milone et al. 2012]{milone12}). Similar analyses, using the two-band photometric
detection method have been conducted on the young (15--30 Myr),
massive cluster NGC 1818, located in the Large Magellanic Cloud (LMC),
producing conflicting results \cite[Elson et al. (1998)]{elson98} and
\cite[De Grijs et al. (2013)]{degrijs13}.  We leveraged our new BINOCS results
(using the cluster M3, with binary and mass info, to show that this
discrepancy within single star cluster is due to differing mass
showing different dynamical ages \cite[(Thompson \& Frinchaboy, 2016)]{thompson16}.  This results has now also been seen
in N-Body simulations \cite[(Geller et al. 2015)]{geller2015}


\vspace*{-0.2 cm}
\begin{thebibliography}{}

\bibitem[Geller \etal\ (2015)]{geller15}
{Geller, A.~M, de Grijs, R., Li, C., Hurley, J.~R., Amari, S., Hoppe, P., Zinner, E., \& Lewis R.S.} 2015,
\textit{ApJ}, 805, 11 

\bibitem[Geller \& Matheiu (2012)]{geller12}
{Geller, A.~M \& Mathieu, R.~D} 2012,
\textit{AJ}, 144, 54 

\bibitem[de Grijs \etal\ (2013)]{degrijs13}
{de Grijs, R.,Li, C., Zheng, Y., Deng, L,. Hu, Y,. Kouwenhoven,
  M.B.N. , Wicker, J.E.} 2013,
\textit{ApJ}, 765, 4 

\bibitem[Milone \etal\ (2012)]{degrijs12}
{Milone, A. P. \etal\ } 2012,
\textit{A\&A}, 540, 16 
%author = {Milone, A. P. and Piotto, G. and Bedin, L. R. and Aparicio, A and Anderson, J. and Sarajedini, A and Marino, A F and Moretti, A and Davies, M B and Chaboyer, B and Dotter, A and Hempel, M and Mar{\'\i}n-Franch, A and Majewski, S and Paust, N E Q and Reid, I N and Rosenberg, A and Siegel, M},

\bibitem[Thompson \& Frinchaboy (2016)]{thompson16}
{Thompson, B.A.\& Frinchaboy, P.M.} 2016,
\textit{AJ}, submitted 


\end{thebibliography}
\end{document}